\documentclass[aps,prl,twocolumn]{revtex4}

\usepackage[pdftex]{graphicx}

\newcommand{\be}{\begin{equation}}
\newcommand{\ee}{\end{equation}}
\newcommand{\bea}{\begin{eqnarray}}
\newcommand{\eea}{\end{eqnarray}}
\newcommand{\HH}{{\cal H}}

\newcommand{\p}{\partial}

\newcommand{\la}{\langle}
\newcommand{\ra}{\rangle}
\newcommand{\lb}{\left[}
\newcommand{\rb}{\right]}
\newcommand{\lp}{\left(}
\newcommand{\rp}{\right)}

\def\nn{\nonumber\\}

\begin{document}

\title{Quench dynamics as a probe of quantum criticality}
%\date{\today}
\author{R. A. Barankov}

\affiliation{Department of Physics, Boston University, Boston MA 02215}

\begin{abstract}

Quantum critical points of many-body systems can be characterized by studying
response of the ground-state wave function to the change of the external
parameter, encoded in the ground-state fidelity susceptibility. This quantity
characterizes the quench dynamics induced by sudden change of the parameter.
In this framework, I analyze scaling relations concerning the probability of
excitation and the excitation energy, with the quench amplitude of this
parameter. These results are illustrated in the case of one-dimensional
sine-Gordon model.

\end{abstract}
\maketitle

Manipulation of quantum many-body systems, a long standing goal of quantum
physics, requires understanding dynamical processes induced by changing
external parameters of the system. The dynamics is particularly interesting
and complex when the system is close to criticality. Recent experiments on
cold gases concentrated on the many-body dynamics at the
Mott-insulator-superfluid transition for lattice bosons~\cite{Greiner02}, the
formation of ferromagnetic order in spinor condensates~\cite{Sadler06}, and
the dynamics of Bose condensation involving spontaneous formation of vortex
pairs in the process of thermal quench~\cite{Weiler08}. These experiments
illustrate that there are several time scales defining the dynamics.

The dynamics induced by parametric perturbation is dictated by the dependence
of the Hamiltonian on parameters and also by the structure of excitations for
a specific system. The relation between the parametric time and the intrinsic
time scales, characterizing the collective dynamics and the relaxation
processes, define the two limiting regimes of adiabatic (slow) and sudden
perturbation.

In the adiabatic limit, where the parametric time is large compared to the
intrinsic time scale, the dynamics of a generic many-body system can be
analyzed using adiabatic perturbation theory. As a result, one finds that upon
crossing a quantum critical point the dynamics is characterized by the rate of
change of the external parameter and by the critical
exponents~\cite{Polkovnikov05,Zurek05}. Specifically, the number of
excitations at large times becomes a universal power-law functions of the
rate. This analysis has been recently extended to the non-linear time
dependence of the control parameter~\cite{Sen08,Barankov08}, and also to the
dynamics of open quantum systems~\cite{Patane08}. An example of a realistic
system where this scaling applies is provided by the problem of adiabatic
loading of strongly-interacting one-dimensional bosons into a commensurate
optical lattice~\cite{Grandi08} that can be mapped to the sine-Gordon model.

The universality of adiabatic dynamics can be compromised in realistic
situations. First, the relaxation time can be comparable or smaller than the
parametric time, especially in higher-dimensional and non-integrable systems,
where thermalization dynamics and also the effects of
environment~\cite{Patane08} complicate the analysis. The details of the
parametric path at a QCP may also become important, and the scaling results
obtained for a single control parameter does not apply.

These limitations are not present for sudden quenches of parameters, i.e. fast
parametric changes. In this case, only the overlap of initial and final states
of the system are important for characterizing the dynamics. The theoretical
analysis of such dynamics has been carried out for specific solvable models.
Specifically, the ordering dynamics in the systems with spontaneously broken
symmetry has been considered in the problem of developing superfluidity after
crossing the Mott insulator-superfluid transition in the Bose-Hubbard
model~\cite{Altman02,Sengupta04}, within the BCS model of superconductivity
where the dynamics is induced by abrupt change of the coupling
constant~\cite{Barankov04,Yuzbashyan05}, and also in the problem of
magnetization ordering in spinor condensates subject to time-dependent
magnetic field~\cite{Lamacraft06}. These examples illustrate sensitivity of
the dynamics to the initial state and final Hamiltonian. Later, the analysis
of correlation functions has been extended to include spatial dependence in
one-dimensional integrable
models~\cite{Calabrese06,Cazalilla06,Cherng06,Manmana07,Kollath07} which
confirmed the physical picture of ‘‘light-cone’’
correlations~\cite{Calabrese06}.

In this work, I study the universal scaling in case of sudden change of a
control parameter, by which the system is driven away from the QCP.
Specifically, I consider a $d$-dimensional quantum system described by a
Hamiltonian $\HH(\lambda)$, where $\lambda$ is an external parameter such that
$\lambda=0$ corresponds to the QCP of the system, and the operator
$V=\p_\lambda\HH$ is a sum of local operators. For $\lambda\neq 0$ the the
system is characterized by finite correlation length, $\xi_{corr}\sim
|\lambda|^{-\nu}$ where $\nu$ is the critical exponent. The dynamical exponent
$z$ relates the critical relaxation time to the value of the control
parameter, $\tau_\xi\sim |\lambda|^{-\nu z}$ (see
Ref.~\cite{Hertz76,Fisher89,Sachdev_book} for details).

The main result of this work is the scaling relations for the probability of
excitation $w_{exc}$ after the control parameter is suddenly changed from the
critical point $\lambda_c=0$ to a finite value $\lambda\ne 0$ away from the
critical point, which is related to the scaling of fidelity susceptibility
$\chi_F$
\bea\label{w_exc_chi_F} 
w_{exc}&\approx & \lambda^2 \chi_F(\lambda)\sim L^d
|\lambda|^{d\nu}\ll 1, \nn
\chi_F&\sim & L^d |\lambda|^{d\nu-2},
\eea
where $L$ is the system size. These results are applicable in the vicinity of
the critical point, $w_{exc}\ll 1$, where the physical properties are fully
characterized by the critical exponents of the transition, and one can rely on
the perturbative analysis. This scaling is closely related to the
recently introduced fidelity susceptibility of the ground-state $\chi_F$, that
measures sensitivity of the ground-state wave-function to the presence of a
quantum critical point~\cite{Zanardi06,Venuti07}.

The main result~(\ref{w_exc_chi_F}) is illustrated in the case of the
sine-Gordon model at the QCP separating the gapless and the gapped phases as a
function of the amplitude of the symmetry-breaking perturbation (see below for
details). In this model, I find that fidelity susceptibility of the
ground-state demonstrates rather peculiar non-analyticity: it diverges or
vanishes at the QCP depending on the coupling constant of the model. This
prediction is in drastic contrast with the known results regarding the
behavior of this quantity at the QCP.

The probability of excitation after a quench
can be obtained using the perturbation theory. Assuming that the final
value of the control parameter $\lambda\ne 0$ is close to the critical value
$\lambda_c=0$, we express the initial state in the basis of the final states
and sum the probabilities of the excitation in this basis:
\be\label{w_ex_chi}
w_{exc}\approx\lambda^2\sum_{n\ne 0}\left|\frac{\la
0,\lambda|V|n,\lambda\ra}{E_0(\lambda)-E_n(\lambda)}\right|^2=\lambda^2\chi_F(\lambda),
\ee
where we used the expression for fidelity susceptibility~\cite{Venuti07} in
terms of the matrix elements of the perturbation operator. Here,
$|0,\lambda\ra$ and $|n,\lambda\ra$ are the ground state and the excited
state, $E_0(\lambda)$ is the ground state energy, $E_n(\lambda)$ is the energy
of the excited state at a given value of $\lambda$.

One obtains the scaling of $w_{exc}$ for a sudden quench employing the scaling
analysis valid close to QCP's~\cite{Hertz76,Fisher89,Sachdev_book}. In this
approach, by using the analogy between the quantum theory at criticality and
the corresponding classical field theory in a higher dimension, one obtains
the scaling dimensions of various quantities. The key observation connecting
the scaling analysis to the quench dynamics is that the scaling transformation
is in fact equivalent to the dynamical problem of quenching the control
parameter from the critical point to a finite value away from the critical
point. According to Ref.~\cite{Venuti07}, the scaling relation for fidelity
susceptibility away from the critical point for the perturbation expressed as
a sum of the local operators, $V=\sum_x v(x)$ takes on the form
\be\label{chi_F_scaling}
\chi_F\sim  L^d|\lambda|^{(2\Delta_V-2z-d)\nu},\nn
\ee
where $z$ is the dynamical exponent, and $\Delta_V$ is the scaling dimension
of $V$.

It turns out that the scaling dimension of the perturbation
operator $V$ can be excluded from this relation. Indeed, we first notice that
the scaling dimension of the operator $V$ is obtained from the
Hellmann-Feynman theorem~\cite{Feynman39} for the derivative of the
ground-state energy
\be
\p_\lambda {\cal E}_{GS}=\la 0\lambda|V|0\lambda\ra/L^d\sim \xi_{corr}^{-\Delta_V}\sim |\lambda|^{\nu\Delta_V},
\ee
where ${\cal E}_{GS}=E_{GS}/L^d$ is the density of the ground-state energy.
Assuming that the scaling dimension of the ground state energy is dominated by
the perturbation operator, one obtains ${\cal E}_{GS}\sim
|\lambda|^{\nu\Delta_V+1}\sim |\lambda|^{2-\alpha}$, where $\alpha$ is the
critical exponent of the ground-state energy~\cite{Fisher89,Sachdev_book}. The
well-known hyperscaling relation~\cite{Fisher89,Sachdev_book} between the
exponents, $2-\alpha=(d+z)\nu$, leads to the identity
\be\label{Delta_V}
\Delta_V=d+z-1/\nu.
\ee
Substituting this relation into Eq.~(\ref{chi_F_scaling}), we arrive at
the main result in Eqs.~(\ref{w_exc_chi_F}). We notice that the
expression for the probability of excitation should be much smaller than
unity, which puts a strict limitation on the final value of the control
parameter after a quench, that depends on the system size. 

The scaling of the excitation energy $E_{exc}=\sum_{n}\lb E_n(\lambda)-E_0(\lambda)\rb w_n$ directly follows from Eq.~(\ref{w_ex_chi}), where $w_{exc}=\sum_{n\ne 0}w_n$. One obtains 
\be
E_{exc}\sim L^d|\lambda|^{(d+z)\nu},
\ee
which is identical to the scaling of the ground-state energy $E_{GS}$. 

Similarly, fidelity susceptibility at the quantum critical point
$\chi_F^c\equiv\chi_F(\lambda=0)$, scales with the system size according to
\be\label{chi_c}
\chi_F^c\sim L^{2/\nu}.
\ee

These results are illustrated in the case of the quantum sine-Gordon model~\cite{Coleman75} defined by the Hamiltonian
($\hbar=1$):
\be\label{Ham_SG}
\HH=\int_0^L dx \lb\Pi^2(x)+(\p_x\varphi)^2-\lambda
\cos\beta\varphi \rb,
\ee
where integration extends over the system size $L$ (the boundary conditions
are not important for the purposes of this work), and the inter-particle
distance defining the large-momentum cut-off of the theory is set to unity.
The canonically conjugate variables $\Pi$ and $\varphi$ obey the standard
commutation relation $\lb \varphi(x), \Pi(x')\rb=i\delta(x-x')$. The coupling
parameter $\beta$ defines two possible physical regimes of the model: the
repulsive one ($4\pi<\beta^2<8\pi$) in which the excitation spectrum of SG
model consists of solitons and antisolitons, and the attractive one
($0<\beta^2<4\pi$) in which the bound states of soliton-antisoliton pairs
(breathers) also form. Individual solitons (antisolitons) have topological
charge $+1$ ($-1$), while breathers have zero topological charge. The spectrum
of quasi-particles is massive at $0<\beta^2<8\pi$ and massless at $\beta^2\ge
8\pi$. Sometimes, it is more convenient to introduce the Luttinger parameter
$K=\beta^2/(4\pi)$, so that the massive regime of the SG model corresponds to
$0\le K<2$. This model describes e.g. the system of interacting lattice bosons
of the particle density commensurate with the lattice
spacing~\cite{Haldane81}.

In the following, I analyze the massive regime and study the critical
properties of the model as one changes the control parameter $\lambda$ close
to the critical point $\lambda_c=0$. This QCP separates the massive regime of
the model at $\lambda\ne 0$ from the massless one at $\lambda_c=0$ that is the
Luttinger liquid~\cite{Haldane81}. The dynamical exponent $z=1$ as it follows
from Eq.~(\ref{Ham_SG}), while the critical index $\nu$ may be extracted from
the expression for the correlation length
\be\label{xi}
\xi_{corr}\sim |\lambda|^{-\nu},\quad
\nu=1/(2-K).
\ee
The spectral gap is identified with the soliton mass $M_s\sim
\xi_{corr}^{-1}\sim |\lambda|^{1/(2-K)}$~\cite{Zamolodchikov95} that enters
the excitation energy of quasi-particles (solitons, antisolitons, and also
breathers at $K<1$) $E_p=\sqrt{p^2+M_s^2}$.

The ground-state fidelity susceptibility of the sine-Gordon
model~(\ref{Ham_SG}) is derived using Eq.~(\ref{w_ex_chi}). For the sake of
clarity, we first discuss this calculation in the repulsive regime ($1<K<2$)
of the model. Since the perturbation operator $V=-\int dx\cos\beta\varphi(x)$
conserves the topological charge and the total momentum of the
quasi-particles, the dominant contribution to the sum over the excited states
in Eq.~(\ref{w_ex_chi}) is provided by the matrix element $\la
0\lambda|V|p,-p;\lambda\ra$ between the ground state $|0\lambda\ra$ and the
excited state that contains a soliton-antisoliton pair
$|n\lambda\ra=|p,-p;\lambda\ra$ with zero total momentum
\be\label{chi_F_sg}
\chi_F\approx L^2\sum_{p}\left|\frac{\la 0\lambda|\cos\beta\varphi|p,-p;\lambda\ra}{2E_p(\lambda)}\right|^2,
\ee
where the summation is over half the relative momentum $p$ of the
quasi-particles, and $E_p=\sqrt{p^2+M_s^2}$ is the excitation energy of
soliton (antisoliton). Since we are interested in the scaling of the fidelity
susceptibility with the control parameter, it is natural to employ the
well-known result~\cite{Coleman75} for the scaling dimension ($K$) of operator
$\cos\beta\varphi$:
\be\label{cos_scaling}
\la 0\lambda|\cos\beta\varphi|0\lambda\ra\sim \xi_{corr}^{-K}\sim |\lambda|^{K/(2-K)},
\ee
where we substituted the correlation length~(\ref{xi}) to obtain the scaling
with the control parameter. Using this relation, it is straightforward to
obtain the scaling of the matrix element involving the soliton-antisoliton
pair, which at small momentum $p\ll \xi_{corr}^{-1}$ assumes the form
\be\label{matrix_massive}
\la 0\lambda|\cos\beta\varphi|p,-p;\lambda\ra\sim \frac{\la 0\lambda|\cos\beta\varphi|0\lambda\ra}{M_sL}\sim \frac{|\lambda|^{K/(2-K)}}{M_sL}.
\ee 
Here, the factor $M_sL$ takes into account the normalization of the excited
state $|p,-p;\lambda\ra$, while another factor comes from the vacuum average.
The large-momentum behavior of the matrix
element~\cite{Barankov_in_preparation} ensures the convergence of the sum in
Eq.~(\ref{chi_F_sg}) for all values of the Luttinger parameter $0<K<2$ in the
massive regime of the model.

Substituting these expressions in Eq.~(\ref{chi_F_sg}), one obtains
\be\label{chi_F_calc}
\chi_F\sim L M_s\frac{|\lambda|^{2K/(2-K)}}{M_s^4}\sim L|\lambda|^{(2K-3)/(2-K)}.
\ee
In this relations the first factor, $LM_s$, comes from the summation over the
states, while another factor comes from the matrix
element~(\ref{matrix_massive}) divided by the excitation energy.

It is straightforward to include $2n$ soliton-antisoliton pairs into this
calculation (the number is even due to conservation of the topological charge,
$n=2,3,..$). Due to conservation of the total momentum, which is zero in the
ground state, one needs to sum over only $2n-1$ momentum states in
Eq.~(\ref{w_ex_chi}), since one of the $2n$ momenta is related to the others
by the equation $p_1+..+p_{2n}=0$. In complete analogy to
Eq.~(\ref{matrix_massive}), we obtain the scaling of the matrix element $|\la
0|\cos\beta\varphi(0)|p_1..p_{2n}\ra|\sim|\lambda|^{K/(2-K)}/(M_sL)^n$. Upon
substitution into the expression for the fideleity, one immediately finds the
same scaling~(\ref{chi_F_calc}) for all $2n$-pair states, as for the one-pair
contribution. We certainly expect, that these multi-pair states provide just a
numerical correction to the dimensionless prefactor of the susceptibility
obtained using the one-pair state~\cite{Barankov_in_preparation}.

One can extend this calculation to the attractive regime at $0<K<1$, where
soliton-antisoliton pairs can form bound states, the so-called breathers. The
essential components of the derivation are not affected by the presence of the
bound states. Specifically, the spectral gap in this case is also proportional
to the soliton mass $M_s\sim |\lambda|^{1/(2-K)}$, and the scaling dimension
of the perturbation operator in Eq.~(\ref{cos_scaling}) is also $K$. Thus, we
conclude that the scaling of the fidelity susceptibility at $0<K<1$ is also
given by Eq.~(\ref{chi_F_calc}).

The scaling relation~(\ref{chi_F_calc}) is in agreement with the general
result~(\ref{w_exc_chi_F}), since $d=1$ and the critical exponent
$\nu=1/(2-K)$. The identity~(\ref{Delta_V}) is also satisfied, as it follows
from the substitution of $d=z=1$, the scaling dimension of the perturbation
operator $\Delta_V=K$, and the critical exponent $\nu$.

Interestingly, upon inspection of Eq.~(\ref{chi_F_calc}), we find that the
character of non-analyticity of the reduced susceptibility $\chi_F/L$ at the
quantum critical point $\lambda_c=0$ depends on the value of the Luttinger
parameter: $\chi_F/L$ diverges at $0<K<3/2$, and it vanishes at $3/2<K<2$ with
$\lambda$.

The divergence and vanishing of the fidelity susceptibility in the
corresponding intervals is accompanied by the similar behavior of the
susceptibility as a function of the system size $L$ at the critical point
$\lambda_c=0$, where the sine-Gordon model reduces to the Luttinger liquid. In
this calculation, we first approximate the fidelity susceptibility by taking
into account only the two-phonon contribution (see Eq.~(\ref{chi_F_sg}) where
$\lambda=0$), and then we study the contribution of multi-phonon states. For
the Luttinger liquid, the vacuum average of the operator $\cos\beta\varphi$,
that defines the behavior of the matrix elements, vanishes in the
thermodynamic limit $L\gg 1$ as dictated by the corresponding scaling
dimension ($K$)
\be\label{cos_scaling_LL}
\la 0|\cos\beta\varphi|0\ra\sim L^{-K}.
\ee
The matrix element between the ground state and the two-phonon at $L\gg 1$ and
at small momenta $p\ll 1$ is given by:
\be\label{matrix_LL}
\la 0|\cos\beta\varphi|p,-p\ra\sim \frac{e^{-|p|}}{L |p|}\la 0|\cos\beta\varphi|0\ra \sim\frac{e^{-|p|}}{L |p|}L^{-K},
\ee
where the factor in the denominator takes into account the normalization of
the momentum states. The matrix element decays exponentially at large momenta
$p\gg 1$ set by the inter-particle distance (unity in our units). Substituting
this result in Eq.~(\ref{chi_F_sg}), where we also use $E_p=p$ for the
excitation energy of the phonons, we find that the sum over the states
diverges at small momenta $p_c\sim L^{-1}$ defined by the system size. The
power-counting leads to the scaling
\be\label{chi_c_calc}
\chi_F/L\sim L^{3-2K}.
\ee

In analogy with the calculation in the massive phase of the sine-Gordon model
($\lambda\ne 0$), one can include the higher-order states in the calculation.
Here, the matrix element between the multi-phonon and the ground state is
given by
\be
|\la 0|\cos\beta\varphi(0)|p_1..p_{n}\ra|\sim \frac{e^{- (|p_1|+..+|p_{n}|)/2}}{\lp L|p_1|..L|p_{n}|\rp^{1/2}}L^{-K}
\ee
It is straightforward to verify that the dominant contribution $\sim L^2$ to
the fidelity susceptibility is provided by the total energy of excitation that
enters the denominator of Eq.~(\ref{w_ex_chi}), which after being combined
with the other factors leads to Eq.~(\ref{chi_c_calc}). This result is in
agreement with Eq.~(\ref{chi_c}). We conclude that the reduced susceptibility
$chi_F^c/L$ diverges at $0<K<3/2$ and vanishes at $3/2<K<2$ in thermodynamic
limit $L\gg 1$.

The analysis of the sine-Gordon summarized in Eqs.~(\ref{chi_F_calc}) and
(\ref{chi_c_calc}) demonstrates that the fidelity susceptibility is not
necessarily a singular function of the control parameter at a QCP, although it
is certainly a non-analytic one. The divergence may be present for some values
of the critical exponent and absent for other ones. Specifically, I find that
in the sine-Gordon model, the susceptibility diverges at a QCP separating the
gapless phase (Luttinger liquid) from the gapped (massive) phase, as a
function of the amplitude of relevant operator $\cos\beta\varphi$, when
$0<K<3/2$ and it vanishes at $3/2<K<2$.

In conclusion, I derived scaling relations for the probability of excitation
and the excitation energy in a system driven away from a QCP, and explored its
relation to fidelity susceptibility of the ground state, as summarized in
Eq.~(\ref{w_exc_chi_F}). These general results are illustrated in the case of
sine-Gordon model, in which I find that fidelity susceptibility diverges at
$0<K<3/2$ and vanishes at $3/2<K<2$ with the control parameter, demonstrating
peculiar type of non-analyticity at the quantum critical point. These results
can be used to probe the properties of QCP's in many-body systems actively
studied in current experiments on cold atoms.

I acknowledge discussions with C.~De~Grandi, V.~Gritsev, and A.~Polkovnikov at
the early stages of this work.

\end{document}